\documentclass{iau} 
\usepackage{graphicx}

\title[Atomic Data for Neutron-Capture Elements] %% give here short title %%
{Atomic Data and Neutron-Capture Element Abundances in Planetary Nebulae}

\author[N.\ C.\ Sterling]
{N.\ C.\ Sterling$^1$}
\affiliation{$^1$Department of Physics, University of West Georgia, 1601 Maple Street, Carrollton, GA 30118, USA \\ email: {\tt nsterlin@westga.edu}} 

\pubyear{2017}
\volume{323}  %% insert here IAU Symposium No.
\jname{Planetary Nebulae: Multi-wavelength probes of stellar and galactic evolution}
\editors{Xiao-Wei Liu, Letizia Stangehllini, and Amanda Karakas, eds.}
\begin{document}

\maketitle

\begin{abstract}

Neutron(\emph{n})-capture elements are produced by \emph{s}-process nucleosynthesis in low- and intermediate-mass AGB stars, and therefore can be enriched in planetary nebulae (PNe).  In the last ten years, \emph{n}-capture elements have been detected in more than 100 PNe in the Milky Way and nearby galaxies.  In some objects, several different \emph{n}-capture elements have been detected, providing valuable constraints to models of AGB nucleosynthesis and evolution.  These detections have motivated theoretical and experimental investigations of the atomic data needed to derive accurate \emph{n}-capture element abundances.  In this review, I discuss the methods and results of these atomic data studies, and their application to abundance determinations in PNe.
\end{abstract}

\firstsection % if your document starts with a section,
\section{Introduction} \label{intro}

Neutron(\emph{n})-capture elements are heavier than zinc (atomic number $Z>30$) and can be produced by slow \emph{n}-capture nucleosynthesis, or the ``\emph{s}-process,'' during  the asymptotic giant branch (AGB) evolutionary stage of planetary nebula (PN) progenitor stars.  The chemical compositions of PNe bear the signatures of AGB nucleosynthesis, and thus can be enriched in \emph{n}-capture elements.  However, detecting the weak emission lines of these species in PNe poses a significant challenge, owing to the low cosmic abundances of \emph{n}-capture elements (less than a few billionths of the H abundance in the Solar System; \cite[Asplund et al.\ 2009]{asplund09}).  The most easily detected trans-iron element ions are those with a small number of ground configuration levels (leading to a few relatively strong emission lines), such as those with $4p^k$ and $5p^k$ valence shells, where $k=1$--5.  Depletion into dust is another factor limiting the detectability of \emph{n}-capture elements, many of which are among the most refractory elements in the Universe (\cite[Lodders 2003]{lodders03}).  Trans-iron elements that are observed in the spectra of AGB stars, such as Sr, Y, Zr, Ba, and La, are likely to be strongly depleted in PNe, making their detection unlikely.

Despite the observational challenges, nebular spectroscopy is a valuable tool for studying \emph{s}-process nucleosynthesis, and reveals information independent of and complementary to stellar spectroscopy.  For example, nebular spectroscopy provides access to elements that cannot be detected in cool giant stars (e.g., the lightest \emph{n}-capture elements and the noble gases Kr and Xe), allowing the production and chemical evolution of these species to be studied.  Moreover, nucleosynthesis is complete in PNe, whose chemical compositions reflect that of the progenitor's AGB envelope during the final few thermal pulses.  Observations of PNe provide constraints to many poorly-understood processes in models of AGB nucleosynthesis, including the efficiency of dredge-up at low envelope masses, convective mixing and mass loss, and to stellar yields (e.g., \cite[Karakas \& Lattanzio 2014]{karakas14}).

Emission lines of \emph{n}-capture elements had been detected in PNe as early as the mid-1970s (\cite[Treffers et al.\ 1976]{treffers76}), but it was not until the pioneering work of \cite[P\'equignot \& Baluteau (1994, hereafter PB94)]{pb94} that features in PN spectra were associated with trans-iron elements. \cite[PB94]{pb94} identified several lines of \emph{n}-capture elements (including Se, Kr, Br, Rb, Xe, and possibly others) in the optical spectrum of the bright PN NGC~7027.  While these identifications garnered some initial skepticism due to the relatively low spectral resolution and possible blending issues for some features, many were confirmed in subsequent studies (e.g., \cite[Sharpee et al.\ 2007]{sharpee07}).

Assessing \emph{n}-capture element abundances required \cite[PB94]{pb94} to make a number of approximations, as the only atomic data available to them were energy levels and transition probabilities (\cite[Bi\'emont \& Hansen 1986a]{biemont86a},\cite[b]{biemont86b}, \cite[1987]{biemont87}; \cite[Bi\'emont et al. 1988]{biemont88}).  For some ions (e.g., Se$^{2+}$ and bromine ions) even the energy levels, compiled by \cite[Moore (1952)]{moore52} from spectroscopic measurements conducted in the early 20th century, were uncertain.  (Recent measurements have greatly improved the accuracy of energy levels for Se$^{2+}$ and Br ions, e.g., \cite[Tauheed et al.\ 2008]{tauheed08}; \cite[Tauheed \& Hala 2012]{tauheed12}; \cite[Jabeen \& Tauheed 2015]{jabeen15}.)  \cite[PB94]{pb94} used collision strengths extrapolated from those of $n=2$ and $n=3$ ions (where $n$ is the principal quantum number of the valence shell) to estimate ionic abundances.  To derive elemental abundances, they used ICFs based on coincidences in the ionization potential ranges with light element ions.  In spite of these approximations, \cite[PB94]{pb94}'s conclusion that NGC~7027 exhibits large \emph{s}-process enrichments has been verified in later studies (\cite[Sharpee et al.\ 2007]{sharpee07}; \cite[Sterling et al.\ 2015]{sterling15}, hereafter SPD15; \cite[Sterling et al.\ 2016]{sterling16}).

Nebular spectroscopy of \emph{n}-capture elements is a field that has developed rapidly over the last 15 years, owing to the identification and detection of emission lines of these elements in several PNe, particularly in the near-infrared.  \cite[Dinerstein (2001)]{dinerstein01} identified two long-anonymous features at 2.199 and 2.287~$\mu$m as fine structure transitions of $[$Kr~III$]$ and $[$Se~IV$]$, respectively; these are the two most widely-detected \emph{n}-capture element lines in PNe (\cite[Sterling \& Dinerstein 2008]{sterling08}).  More recently, \cite[Sterling et al.\ (2016)]{sterling16} discovered $[$Ge~VI$]$, $[$Rb~IV$]$, and $[$Cd~IV$]$ lines in the near-IR spectra of two PNe, while $J$~band transitions of $[$Se~III$]$ and $[$Kr~VI$]$ have been identified by \cite[Sterling et al.\ (2017)]{sterling17}.  To date, \emph{n}-capture elements have been detected in more than 100 PNe in the Milky Way (e.g., \cite[Sharpee et al.\ 2007]{sharpee07}; \cite[Sterling \& Dinerstein 2008]{sterling08}; \cite[Garc{\'{i}}a-Rojas et al.\ 2015]{garcia-rojas15}) and nearby galaxies (\cite[Otsuka et al.\ 2011]{otsuka11}; \cite[Mashburn et al.\ 2016]{mashburn16}).

These detections have motivated laboratory astrophysics efforts to determine the atomic data needed for accurate \emph{n}-capture element abundance determinations in PNe.  The atomic data fall into two categories: those needed for ionic abundances, and those for elemental abundances.  To derive ionic abundances, transition probabilities and effective collision strengths for electron-impact excitation are required.  Shortly after the publication of \cite[PB94]{pb94}, \cite[Bi\'emont et al.\ (1995)]{biemont95} calculated transition probabilities for $5p^k$ ions including Xe, while \cite[Sch\"oning (1997)]{schoning97} and \cite[Sch\"oning \& Butler (1998)]{sb98} computed effective collision strengths for various Kr, Xe, and Ba ions.  For elemental abundance determinations, ionization correction factors (ICFs) must be used to account for the abundances of unobserved ions.  The most reliable ICF prescriptions are determined from photoionization models (\cite[Delgado-Inglada et al.\ 2014]{delgado-inglada14}; \cite[SPD15]{sterling15}; see also the review of C.\ Morisset in this volume) that compute the ionization equilibrium of each element.  Such models rely on the availability of accurate photoionization (PI) cross sections and rate coefficients for radiative recombination (RR), dielectronic recombination (DR), and charge transfer (CT) -- none of which had been determined for \emph{n}-capture element ions until recently.

In this review, I focus on atomic data developments for \emph{n}-capture elements in the last 10 years, along with their application to observations.

\section{Theoretical Methods and Results} \label{theory}

Due to the atomic data needs for several ions over a large range of energies and temperatures, the majority of atomic data determinations for photoionization and spectral modeling are theoretical.  The initial step in these calculations is modeling the structure of the ions of interest.  This yields energy levels, transition probabilities, and radial orbital wave functions.  At present, the most widely-used codes for atomic structure calculations are AUTOSTRUCTURE (\cite[Badnell 2011]{badnell11}) and GRASP (e.g., \cite[J\"onsson et al.\ 2007]{jonsson07}).  AUTOSTRUCTURE includes first-order relativistic corrections using Breit-Pauli formalism, while GRASP is based on the fully-relativistic Dirac-Fock method.  Both approaches account for configuration interaction (mixing between electronic states) and allow for semi-empirical corrections to observed energy levels.

The orbital wave functions obtained in the structure calculations are essential ingredients to calculations of photoionization, electron-impact excitation, and recombination cross sections.  Two general methods are used in such scattering calculations.  The distorted wave (DW) method uses distorted wave functions to compute bound-free interactions.  No correlation effects between bound and continuum states are taken into account, and hence the cross sections do not include resonances.  However, radiative and autoionizing rates for levels above the ionization threshold can be calculated using the isolated resonance approximation, allowing DR rate coefficients to be computed with this method.  DW calculations are computationally efficient and thus are a pragmatic approach for producing atomic data for large numbers of systems.  Another benefit of the computational efficiency is that uncertainties in the resulting atomic data can be estimated (e.g., \cite[Sterling \& Witthoeft 2011]{sterling11b}), for example by using different configuration expansions.  The DW method is preferred for computing RR rate coefficients (\cite[Badnell 2006]{badnell06b}), and has been used to calculate DR rate coefficients for numerous ions (\cite[Badnell et al.\ 2003]{badnell03}; \cite[Abdel-Naby et al. 2012]{abdel-naby12}, and references therein).  For processes in which resonances play a crucial role and the isolated resonance approximation becomes inaccurate, such as photoionization and collisional excitation for low-charge ions, the R-matrix method is preferred.  R-matrix calculations use the close-coupling approximation, which includes all correlation effects between bound and continuum states within the limits of the configuration expansion.  The R-matrix method produces the most accurate data for resonant processes, but is significantly more computationally intensive than DW calculations.

PI, RR, and DR calculations for Se and Kr ions (up to five times ionized) have been computed in the DW approximation (\cite[Sterling \& Witthoeft 2011]{sterling11b}; \cite[Sterling 2011]{sterling11c}).  We have also completed calculations of these data for Br, Rb, and Xe ions, which are being prepared for publication.  While the PI cross sections do not have accurate resonance energies, the direct cross sections (corresponding to direct ionization of a valence electron) agree well with experimental measurements (\S\ref{exp}).

\begin{figure}[t]
% \vspace*{-2.0 cm}
\begin{center}
 \includegraphics[width=5.3in]{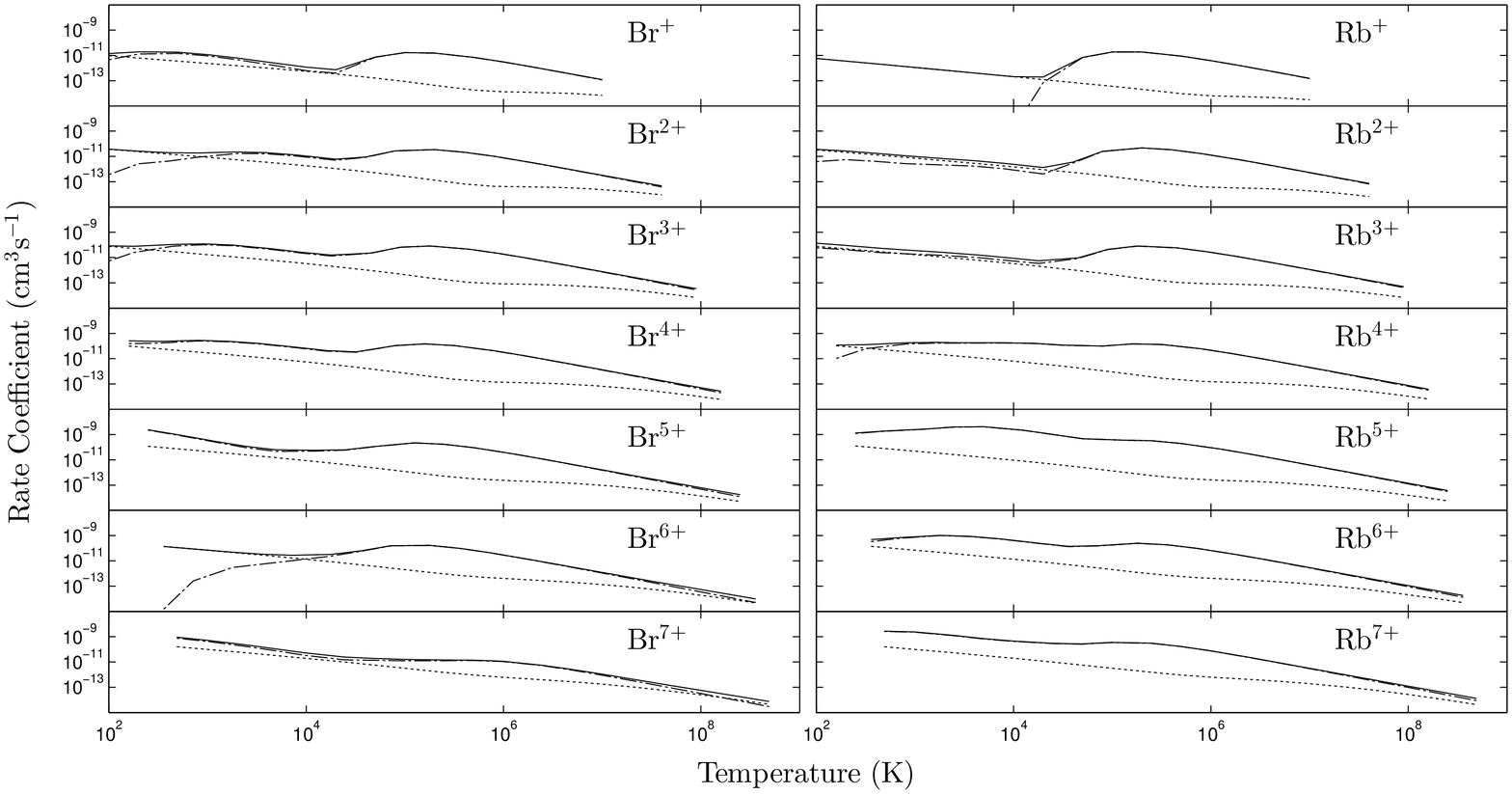} 
% \vspace*{-1.0 cm}
 \caption{Comparison of RR (dotted lines), DR (dot-dashed lines), and total recombination (solid lines) rate coefficients for Br and Rb ions.}
   \label{brrb}
\end{center}
\end{figure}

We estimated uncertainties for all processes, with typical error bars of 30--50\% for PI, 10--30\% for RR, and factors of 2--10 for DR.  The large uncertainties for DR rate coefficients are important, since DR is the dominant recombination mechanism for low-charge \emph{n}-capture element ions at the ``low'' temperatures of PNe (as illustrated for Br and Rb ions in Figure~\ref{brrb}).  Low-temperature DR rate coefficients of heavy elements cannot be precisely calculated even with the most sophisticated theoretical methods, due to the unknown energies of autoionizing states that lie just above the ionization threshold.  This is an important uncertainty in ionization equilibrium solutions for all elements beyond the second row of the periodic table.

R-matrix PI cross section calculations have been performed for selected \emph{n}-capture element ions: Se$^+$ (\cite[McLaughlin \& Ballance 2012a]{mclaughlin12a}), Se$^{2+}$ (\cite[Macaluso et al.\ 2015]{macaluso15}), Kr$^+$ and Xe$^+$ (\cite[McLaughlin \& Ballance 2012b]{mclaughlin12b}), and Rb$^{2+}$ (\cite[Macaluso et al.\ 2016]{macaluso16}).  These data agree well with experimental measurements, in terms of both the absolute cross sections and the resonance energies.  R-matrix PI calculations are underway for Br and Rb ions.

The recent discovery of near-IR \emph{n}-capture element transitions has spurred R-matrix calculations of effective collision strengths.  In addition to the work of Sch\"oning and Butler for Kr, Xe, and Ba (see \S\ref{intro}), collision strengths have been computed for Se$^{3+}$ (K.\ Butler 2007, private communication), Rb$^{3+}$ (\cite[Sterling et al.\ 2016]{sterling16}), and Se$^{2+}$ and Kr$^{5+}$ (\cite[Sterling et al.\ 2017]{sterling17}).  The relative intensities for different Rb$^{3+}$ and Se$^{2+}$ lines agree well with radiative-collisional models that incorporate the new collision strengths, supporting the accuracy of these calculations.  No such comparison is possible for Se$^{3+}$ and Kr$^{5+}$, whose $4s^24p$ ground configurations lead to a single transition.  Additional collision strength calculations for Br and Rb ions detected in PNe are in progress.

Charge transfer recombination with neutral H atoms cannot be treated using the codes described above, because it is a quasi-molecular process.  Using the Landau-Zener method (\cite[Butler \& Dalgarno 1980]{bd80}; \cite[Janev et al.\ 1983]{janev83}) for multiply-charged atoms and the Demkov approximation (\cite[Demkov 1964]{demkov64}; \cite[Swartz 1994]{swartz94}) for singly-ionized species, \cite[Sterling \& Stancil (2011)]{sterling11d} computed CT rate coefficients for Ge, Se, Br, Kr, Rb, and Xe ions up to five times ionized.  These calculations are accurate to within a factor of three for systems with large rate coefficients (\cite[Butler \& Dalgarno 1980]{bd80}), and are less accurate for those with smaller rate coefficients (although RR and DR typically dominate recombination for those ions).  However, state-of-the-art CT calculations such as the quantum mechanical molecular-orbital close-coupling (QMOCC) method (e.g., \cite[Wang et al.\ 2004]{wang04}) are computationally intensive, and must be targeted for ions in which uncertainties in CT rate coefficients most significantly affect abundance determinations.

\section{Experimental Methods and Results} \label{exp}

Although the majority of atomic data for \emph{n}-capture elements are determined theoretically, it is critical to benchmark the results against experimental measurements when possible, due to the complexity of these many-electron systems.  PI cross sections can be measured via the merged beams technique (\cite[Lyon et al.\ 1986]{lyon86}) at third-generation synchrotron radiation facilities.  Experimental setups for measuring PI cross sections that are in current operation include the PIPE apparatus (\cite[Rics\'oka et al.\ 2009]{ricsoka09}) at PETRA~III in Germany and the MAIA apparatus at SOLEIL in France (\cite[Bizau et al.\ 2016]{bizau16}).  Several of the results below were obtained with the now-defunct IPB apparatus (\cite[Covington et al.\ 2002]{covington02}) at the Advanced Light Source (ALS; USA) and a merged-beams setup on the Miyake undulator beamline at ASTRID (Denmark).

In the merged-beams method, atomic ion beams are produced with a plasma source, and the charge-to-mass ratio of the desired ion is selected with a dipole analyzing magnet.  A series of electrostatic lenses and steering plates are used to focus the ion beam, which is merged with a counter-propagating beam of UV/soft X-ray photons.  The resulting photoions are directed to a single-particle detector by a demerging magnet.  Using monochromatic gratings to step the photon energy, photoionization yields can be measured as a function of energy.   The relative cross sections are placed on an absolute scale at discrete energies by measuring the photoion yields produced in a well-defined volume, which entails characterizing the beam overlap with scanning slits.  The disadvantage of the merged beams method is that the ion beams are not purely in the ground state (unless an ion trap is used; e.g., \cite[Bizau et al.\ 2011]{bizau11}).  The measured cross sections are linear combinations of those from the ground and metastable states.  Therefore R-matrix calculations are needed to disentangle the contribution of each ground configuration state to the measured cross sections.

\begin{figure}[t]
% \vspace*{-2.0 cm}
\begin{center}
 \includegraphics[width=3.4in]{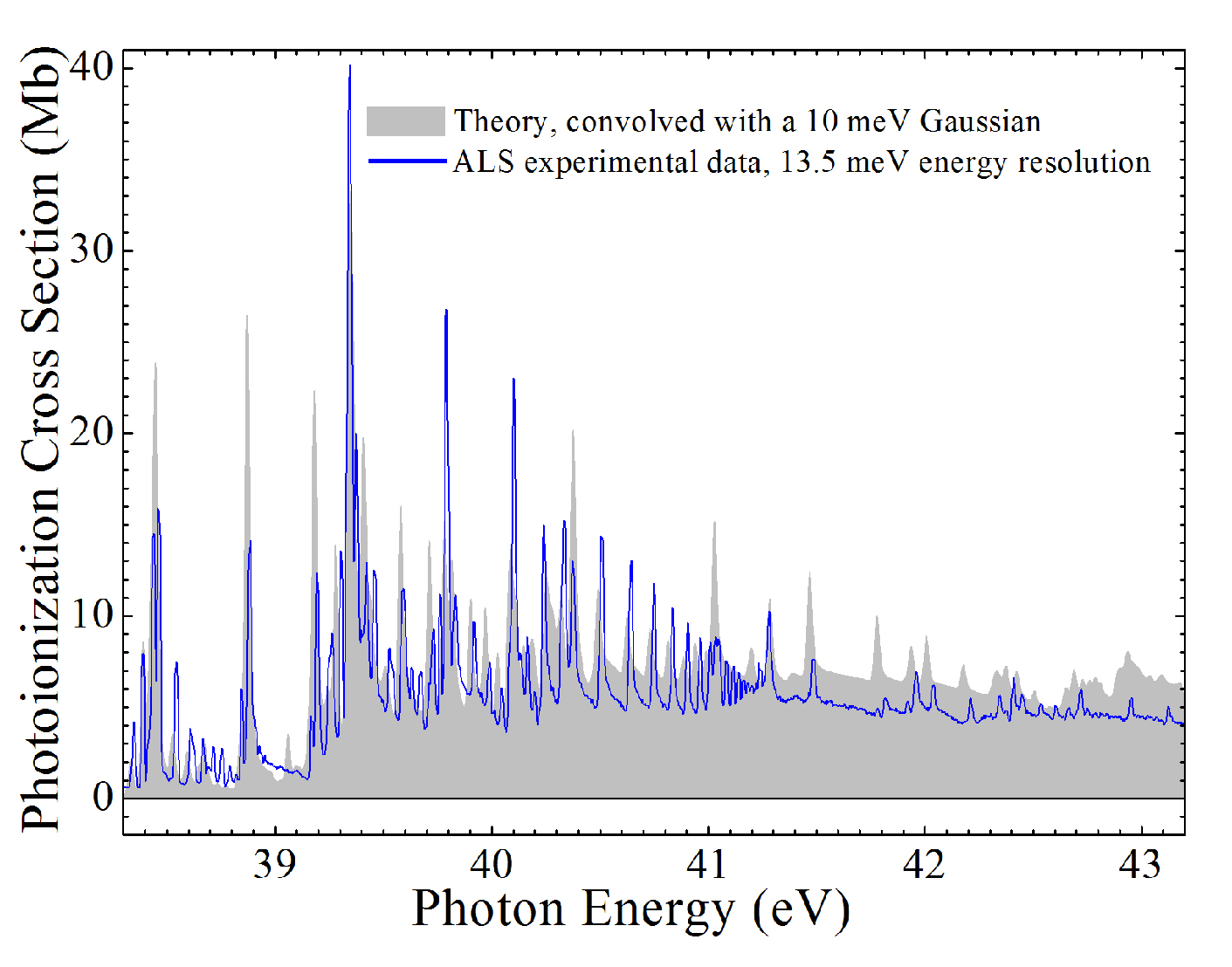} 
% \vspace*{-1.0 cm}
 \caption{Comparison of theoretical and experimental Rb$^{2+}$ photoionization cross section.  Figure adapted from \cite[Macaluso et al.\ (2016)]{macaluso16}.}
   \label{rb2pi}
\end{center}
\end{figure}

Experimental PI cross section measurements have been completed for ions X$^{+q}$ ($q=1$--5) of Se (\cite[Sterling et al.\ 2011]{sterling11a}; \cite[Esteves et al.\ 2012]{esteves12}; \cite[Macaluso et al.\ 2015]{macaluso15}), Br, Kr (\cite[Lu et al.\ 2006a]{lu06a},\cite[b]{lu06b}; \cite[Bizau et al.\ 2011]{bizau11}), Rb (\cite[Macaluso et al.\ 2016]{macaluso16}), and Xe (\cite[Bizau et al.\ 2006]{bizau06}, \cite[2011]{bizau11}).  The experimental data for Br and Rb ions are being analyzed (and R-matrix calculations underway), with the exception of the recently-published results for Rb$^{2+}$ (Figure~\ref{rb2pi}).

Experimental measurements of other atomic processes have not been conducted for \emph{n}-capture elements.  However, the recently constructed Cryogenic Storage Ring (\cite[von~Hahn et al.\ 2016]{vonhahn16}) can be used to measure total recombination rate coefficients (for both RR and DR) of low charge-to-mass species such as the \emph{n}-capture element ions detected in PNe.  Given the large uncertainties in theoretical DR rate coefficients, such measurements would be particularly valuable for improving the accuracy of \emph{n}-capture element abundance determinations.

\section{Applying the New Atomic Data to Observations}

Applications to observational data provide the ultimate test of the new atomic data.  PNe have long served as ``laboratories'' for assessing the accuracy of atomic data.  Atomic data for light elements have been improved through observations paired with increasingly sophisticated theoretical and experimental techniques, and one can expect a similar scenario to unfold for \emph{n}-capture elements.

\cite[SPD15]{sterling15} added Se and Kr to the atomic database of Cloudy (\cite[Ferland et al.\ 2013]{ferland13}) in order to determine ICFs for these elements.  They modeled 15 PNe, with input parameters such as central star temperature and luminosity, nebular density, outer radius, and elemental abundances determined via optimization routines that provided the best fit to the observed emission line spectrum of each object.  These 15 PNe were chosen because multiple Kr ions had been detected in their optical and near-IR spectra.  It was found that Cloudy models were unable to reproduce the ionization balance of Kr in these PNe, with the Kr$^{3+}$/Kr$^{2+}$ ratio systematically lower in the models than observed, regardless of the ionization level of the PN.  To rectify this discrepancy, the PI cross sections of Kr$^{2+}$, Kr$^{3+}$, and Kr$^{4+}$ were adjusted within their estimated uncertainties, and the DR rate coefficients of these ions by amounts slightly larger than the stated error bars.  These empirical adjustments should be viewed as temporary measures that highlight the need for further laboratory astrophysics studies (e.g., experimental DR measurements) of the ions involved.  More physically realistic 3D photoionization models, that take into account morphological and density structures, would also provide valuable tests of 1D model results for the Kr ionization balance.

\cite[SPD15]{sterling15} could not perform a similar analysis for the ionization equilibrium of Se, since only $[$Se~IV$]$ had been detected in PNe at the time (with the possible exception of $[$Se~III$]$~8855.27~\AA\footnote{Air wavelengths for $[$Se~III$]$ computed from the energy levels of \cite[Tauheed \& Hala (2012)]{tauheed12}}, which is significantly blended with He~I~$\lambda$8854.20).  However, observations of the uncontaminated $[$Se~III$]$~1.0991~$\mu$m line identified by \cite[Sterling et al.\ (2017)]{sterling17} can be used to test the atomic data and ICF schemes for Se.

\cite[SPD15]{sterling15} computed large grids of models that span the range of nebular densities, metallicities, and central star temperatures and luminosities of PNe in the Milky Way and nearby galaxies.  Correlations between the fractional abundances of Kr/Se ions and those of routinely detected ions of O, S, Cl, and Ar were fitted with analytical functions, the inverses of which serve as ICFs.  These ICFs were applied to a sample of 120 PNe (\cite[Sterling \& Dinerstein 2008]{sterling08}) and produced Se and Kr abundances 0.1--0.3~dex lower than previous estimates.  However, the new abundances did not affect the conclusions of \cite[Sterling \& Dinerstein (2008)]{sterling08} or the correlations they found between Se and Kr abundances and other nebular and stellar properties.  Recent results from deep, high-resolution optical spectra of PNe (\cite[Garc\'ia-Rojas et al.\ 2015]{garcia-rojas15}; \cite[Madonna et al.\ 2017]{madonna17}) show good agreement between the Kr abundances derived with the various ICF prescriptions of \cite[SPD15]{sterling15}.  New Kr ICFs that include Kr$^{5+}$ have been computed by \cite[Sterling et al.\ (2017)]{sterling17}.

Efforts are underway to add Br, Rb, and Xe to the atomic database of Cloudy.  The atomic data for Xe will be tested as for Kr, via comparisons of modeled and observed line intensities for different Xe ions (unfortunately, multiple Br and Rb ions have been detected in too few PNe thus far to rigorously test their atomic data).  ICFs for these elements will be derived from grids of models.

\section{Future Work}

Atomic physics forms the foundation on which spectroscopy is built.  The range of energies and temperatures for which atomic data must be determined necessitates a predominantly theoretical approach.  However, all theoretical treatments of several-electron ions are approximations, and therefore observations are needed to assess the accuracy of the atomic data and ICF formulae.  Detections of multiple Br and Rb ions in the optical and near-IR spectra of PNe are sorely needed to test atomic data for these elements.  Observations of the newly-identified $[$Se~III$]$~1.0991~$\mu$m line in PNe with a range of ionization levels are required to test the Se ICFs computed by \cite[SPD15]{sterling15}.

The accuracy of ionization equilibrium solutions (and hence ICFs) can also be improved via further experimental and theoretical atomic data investigations.  Specifically, the large uncertainties of theoretical DR rate coefficients for \emph{n}-capture elements can be addressed with experimental measurements at the new Cryogenic Storage Ring facility in Germany, and more accurate CT rate coefficient calculations (e.g. with the QMOCC method) may be needed for specific ions.  Monte Carlo routines are being developed to test the sensitivity of emission line strengths (and hence abundances) to uncertainties in PI, RR, DR, and CT atomic data for individual ions.  The results will help to identify the ions and atomic processes that most urgently require further investigation.

The number of \emph{n}-capture element detections and new identifications will continue to grow, which will lead to new atomic data needs.  The synthesis of observations, atomic physics, and photoionization modeling is essential to continue honing nebular spectroscopy into an effective tool for studying \emph{s}-process enrichments in PNe and hence AGB nucleosynthesis and evolution.

\section{Acknowledgments}

I am grateful to M.\ A.\ Bautista for his helpful comments on \S\ref{theory}, and J.-M.\ Bizau and R.\ C.\ Bilodeau for sending information about currently operating instrumentation for experimental PI measurements.  I also thank my (undergraduate) students who have contributed to the research described herein: John Harrison, Austin Kerlin, Amanda Mashburn, Cameroun Sherrard, and Courteney Spencer.  I acknowledge support from NSF grant AST-1412928.

%\bibliographystyle{mn2e}
%\bibliography{review_iau.bib}

\end{document}